\documentclass{iaus}
\usepackage{graphicx}
\usepackage{wasysym}

\title[Stellar convection, pulsation and semiconvection] %% give here short title %%
{Simulations of stellar convection, pulsation and semiconvection}

\author[Herbert J. Muthsam et al.]   %% give here short author list %%
{Herbert J. Muthsam$^1$, Friedrich Kupka$^1$, Eva Mundprecht$^1$, Florian Zaussinger$^{2,1}$ ,
 Hannes Grimm-Strele$^1$ \& Natalie Happenhofer $^1$}

\affiliation{$^1$Faculty of Mathematics, University of Vienna \\ Nordbergstrasse 15,
A-1090 Vienna, Austria \\ email: {\tt herbert.muthsam@univie.ac.at} \\[\affilskip]
$^2$Max Planck Institute for Astrophysics, \\ 
Karl Schwarzschildstrasse 1, D-85741 Garching, Germany}

\pubyear{2010}
\volume{IAU Symposium No. 271}  %% insert here IAU Symposium No.
\pagerange{000--007}
\jname{Astrophysical Dynamics: From Stars to Galaxies}
\editors{A.C. Editor, B.D. Editor \& C.E. Editor, eds.}
\begin{document}

\maketitle

\begin{abstract}
We report on modelling in stellar astrophysics with the ANTARES code. First, we describe properties of 
turbulence in solar granulation as seen in high-resolution calculations. Then, we turn to the first 
2D model of pulsation-convection interaction in a cepheid. We discuss properties of the 
outer and the HeII ionization zone. Thirdly, we report on our work regarding 
models of semiconvection in the context of stellar physics.
\end{abstract}

\firstsection % if your document starts with a section,
\section{Introduction}
In this paper we present applications of the ANTARES code to various 
problems in stellar physics. ANTARES is a radiation-hydrodynamics code. Its 
core functionality has recently been described by \cite{muth10}. Using this core 
functionality, we first discuss properties of turbulence in \textit{solar granulation}, studied in extremely 
high resolution in 3D.

We have in the meanwhile extended the core functionality of ANTARES, and the applications discussed in the
next two chapters make use of these extensions. So, we have included a spherical and optionally radially 
moving grid for the study of the interaction of radial stellar pulsation with convection. We present a first description of the 
pulsation-convection interaction of a 2D 
model of a \textit{cepheid} with realistic equation of state and grey radiative transfer. For the 
importance of studying the pulsation-convection interactions consult, for example, the reviews 
\cite{buch97} and \cite{buch09}. Consider that since the pioneering work due to Christy, Cox and 
Kippenhahn in the 1960's improvement in nonlinear stellar pulsation modelling has 
basically been due to better microphysics and to some extent to improvement in numerics. Convection has, however, 
(practically) always been included by a mixing-length type approach where already the basic form of the 
equations may be prone to some doubt and where additionally a number of parameters needs to be set in a way which 
can barely considered to be really convincing. Problems probably stemming from such shortcomings are 
discussed, e.g., in \cite{buch97}.
For a recent discussion of a specific uncertainty which arises from 
traditional convection modelling in pulsating stars (namely, excitation of double mode pulsations or 
lack thereof) see  \cite{smol09}.

Subsequently, we turn our attention towards modelling of \textit{semiconvection} in the 
context of stellar physics. Although it is textbook knowledge that a faithful description of semiconvection 
is highly important for properly dealing with the late stages of stellar evolution, but little work has been undertaken 
with regard to multidimensional modelling in the stellar context. The only major numerical studies are due to 
\cite{merry95} and \cite{biello01}. See also the results due to \cite{basc07}. We present here our first 2D results discussing, in particular, the properties of diffusive interfaces. These results mainly refer to 
Boussinesq models, although we report also on compressible models and discuss their connection with the Boussinesq case. 
For enabling a more extensive investigation of compressible models it is necessary to implement a method 
which allows large time-steps, restricted by the macroscopic rather than the sound velocity. In this connection, 
an implementation of the method of \cite{kwatra09} is in an advanced stage of development.

\section{High-resolution solar granulation simulations}

\begin{figure}[b]
% \vspace*{-2.0 cm}
\begin{center}
 \includegraphics[width=3.4in]{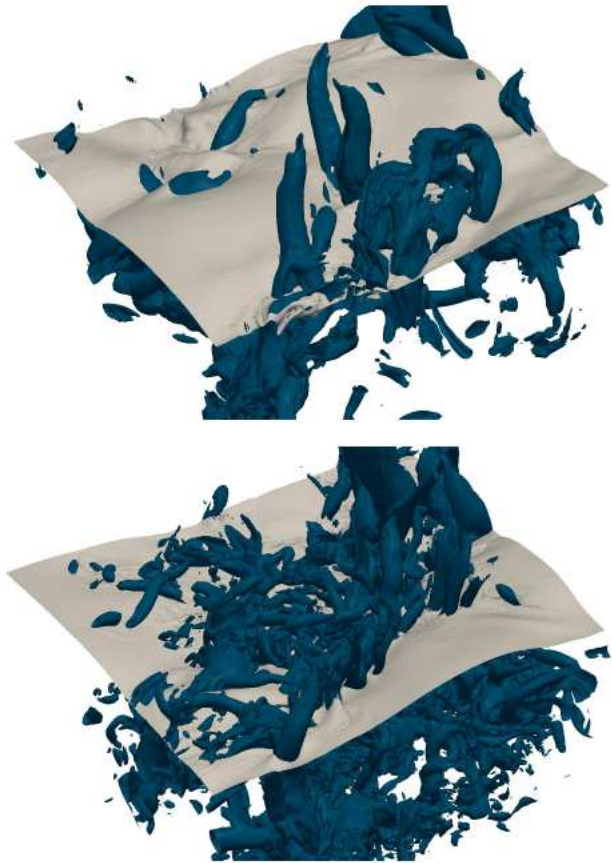} 
% \vspace*{-1.0 cm}
 \caption{Solar granulation at two instances of time. Horizontal extent of the 
 domain shown: $1\mbox{ Mm}$. The grey surface is an isosurface for $T=6000\mbox{ K}$. 
 The vortex tubes are made visible by plotting isosurface for a (large) value of the modulus of the 
 gradient of a suitably normalized pressure. A comparison of the upper and lower figure shows the effect 
 of increased degree of turbulence when applying higher resolution.}
   \label{fig:X}
\end{center}
\end{figure}

In order to investigate the turbulent state of solar granulation we have performed hydrodynamic 
simulations with extremely high resolution. While they are basically similar to the 
ones described in \cite{muth10}, they differ in that the highest resolution is achieved by  making use of \textit{two} grid refinement 
zones (stacked within each other) instead of one, and the resolution in the smaller one 
is considerably finer than in the old run. In addition, the old run referred to an exploding granule, 
whereas we concentrate here, in the highest resolution, on a strong downdraft surrounded by normal granules in 
order to figure out to what extent our previous results apply here as well.

More specifically, the basic computational domain spans now $3.68\times 6\times 6\mbox{ Mm}^3$ (the first 
coordinate is the vertical) with a cell size of $10.8\times 22.2\times 22.2\mbox{ km}^3$. The first grid 
refinement zone has an extension of 
$1.9\times 2.5\times 2.8\mbox{ Mm}^3$ and a grid spacing of 
$5.4\times 7.4\times 7.4\mbox{ km}^3$. 
The second and therefore 
finest grid refinement zone contains the downdraft mentioned above and parts of the surrounding granules. It has an extent of $1\times 1.2\times 1.2 \mbox{ Mm}^3$ with a cell size of $2.7\times 3.7\times 3.7\mbox{ km}^3$.

Using the high-resolution calculations, we want to discuss the role of resolution in the representation of the
turbulent state of the solar granulation layer. In earlier simulations of an exploding granule \cite{muth10} a host 
of vortex tubes of small diameter has shown up, located preferentially near the granular lanes and 
the downdrafts. The questions 
popping up are now whether a similar large number of vortex tube shows up even in more normal granulation 
and whether these previous calculations were fully resolved; they used a grid spacing of 
$7.1\times 9.8\times 9.8\mbox{ km}^3$, i.e. worse than even our present \textit{first} refinement.

In order to answer the question at what resolution the turbulent field of solar granulation is 
(possibly) fully resolved, we compare in figure \ref{fig:X} two snapshots of our highest resolution subdomain.
The first figure represents the state shortly after we have turned on the second grid refinement and thus still 
have the effective resolution due to the first refinement, whereas the next snapshot is taken $153$ seconds later.
Due to increased resolution, the number and intensity of the vortex tubes has drastically increased. A closer study
of the properties of this special sort of turbulence encountered mainly in granular downdrafts is under way.

\section{Cepheid pulsation and convection: a 2D model}

\textit{Initial condition and computational domain.}
We model a cepheid with an
effective temperature of $5125\mbox{K}$, luminosity $L=912.797L_{{\scriptscriptstyle \bigodot}}$,
mass $M=5.0M_{{\scriptscriptstyle \bigodot}}$, hydrogen content
$X=0.7$ and metallicity $Z=0.01$. Our computational domain reaches from
$4000\mbox{ K}$ to $320.000\mbox{ K}$, thus the outer $42\mbox{\%}$ of the star are computed.

This domain is equipped with a polar grid. In radial direction $N_{x}=$
510 grid points plus 4 ghost cells at each boundary are used. The
r-range covers $r\in\left[r_{top},r_{bot}\right],$ where $r_{bot}$
is fixed at $15.5\mbox{ Gm}$ and $r_{top}$ varies with time from about $26.2\mbox{ Gm}$ to 
$27.4\mbox{ Gm}$. The
grid is stretched in radial direction by a factor $q$. The mesh sizes
are $\Delta r_{i+1}$=$q\Delta r_{i}$ varying from $\Delta r_{0}=0.046\mbox{ Mm}$
at the top to $\Delta r_{N_{x}-1}=12\mbox{ Mm}$ at the bottom. In angular
direction there are 800 grid points, the distance $r_{i}\Delta\varphi$
between two adjacent gridpoints in the H-ionisation zone is $5.8\mbox{ Mm}$ and 
$5.6\mbox{ Mm}$ in
the He-ionisation zone. The corresponding aspect ratios are $1:2.5$  and $1:0.6$.

\textit{Fluxes and pulsation in the HeII-ionisation zone.}
The work integral can be decomposed into 
an average part 
$PdV_{0}=\partial_r u_{0}(p/\rho)$  %%$PdV_{0}=\frac{\partial u_{0}}{\partial r}\frac{p}{\rho}$.a perturbational part of
and a perturbational part defined as 
$PdV_{pert}=\partial_r u_{r}^{'}(p/\rho)$.  %%$PdV_{pert}=\frac{\partial u_{r}^{'}}{\partial r}\frac{p}{\rho}$
Here, $u_{0}=\bar{I_{r}}/\bar{\rho}$ and $u_{r}'=u_{r}-u_{0}$. 
$\vec{I}=\left(I_{r},I_{\varphi}\right)$ is the momentum vector
and $\vec{u}=\left(u_{r},u_{\varphi}\right)$ the velocity. $\bar{I_{r}}$
denotes the horizontal average of the radial momentum component.

\begin{figure}
\includegraphics[scale=0.9]{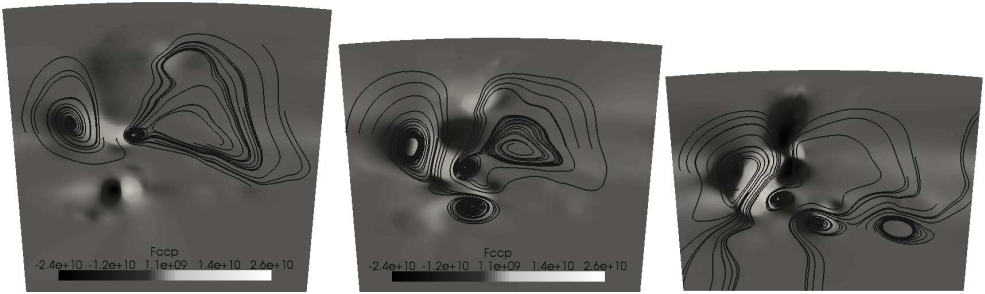}\caption{Convective flux in the course of  contraction (HeII convection zone)}
\label{fig:fconv_puls}
\end{figure}
\begin{figure}
\begin{center}
\includegraphics{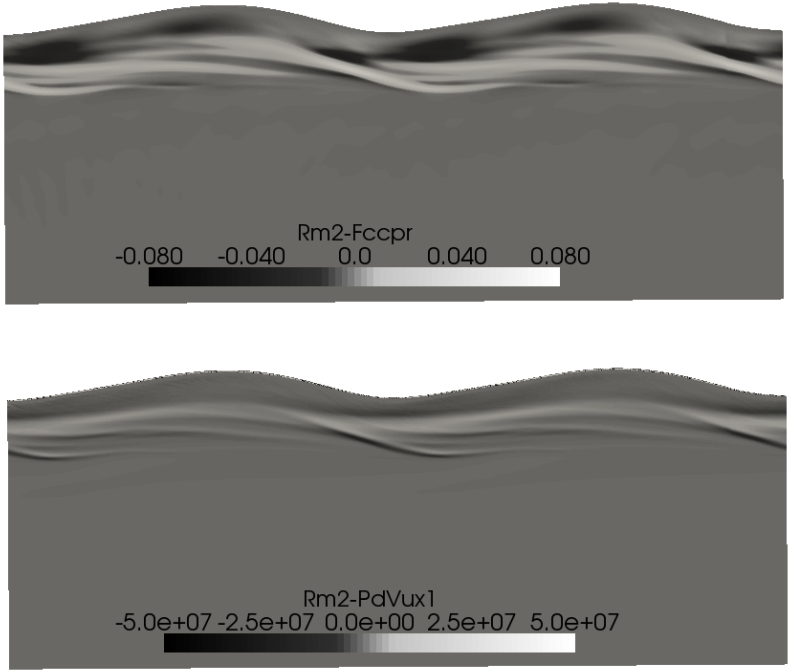}\caption{Horizontal averages of convective flux (upper figure) and work integral 
(lower figure)}
\label{fig:pulsprops}
\end{center}
\end{figure}

While $PdV_{pert}$ varies locally from $-6.e8$ to $+5.e8$ in the HeII convection zone and is
greater than $PdV_{0}\in\left[-8.e7,+5.e7\right]$, the horizontal
average is considerably smaller, $\overline{PdV_{pert}}\in\left[-9.e6,+7.e6\right]$,
compared to $\overline{PdV_{0}}\approx PdV_{0}$.

The averaged convective flux varies from $-3\%$ to $+8\%$ of the input energy flux at the
lower boundary, the mean
value during one period being $+4\%$. Locally values of up to $1.4e11$ corresponding to $30\%$ can be reached. The averaged kinetic flux varies
from $-3\%$ to $+0.2\%$ the mean value is $-2\%$. The data are from 12 consecutive periods and the same flux pattern can be observed in each though the extent may vary. In the expanded state one observes two convection centers: a new one forming at the top and an older one which is the remainder of the downwards moving and slowly disapearing plume of the previous period. At contraction the two centers are very close to each other. These centers of convective activity lead also
to the two (occasionally more)  stripes situated atop of each other and visible, for example, in the quantities
depicted in figure \ref{fig:pulsprops}. The centers of activity can also directly be seen in 
figure \ref{fig:fconv_puls}.

\textit{Remarks on the H-ionisation zone.}
\begin{figure}
\begin{center}
\includegraphics[scale=0.75]{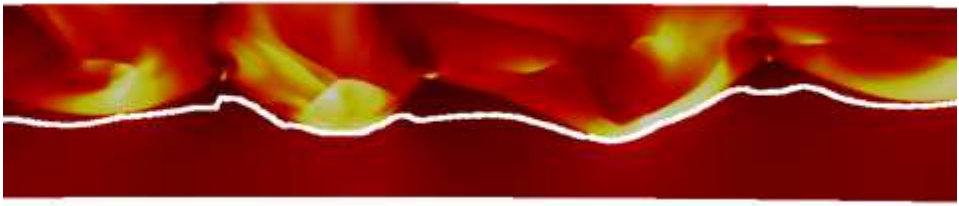}
\caption{Radial momentum in H-ionisation zone. The bright line denotes the optical depth $\tau=100$}
\label{fig:uppercz}
\end{center}
\end{figure}
The upper boundary of our computational domain is situated at an optical deph of a few thousandths. Figure
\ref{fig:uppercz} shows a part (in $\phi-$direction) of the upper zone. Color coding is for momentum densities 
(bright resp. red is downwards). The convective motions are very vigorous. We have observed Mach numbers for 
inflow of up to 4 and for outflow of up to 3. Remarkably, the lower boundary of the downdrafts often 
coincides the line of maximum temperature gradient (just above the bright line in figure \ref{fig:uppercz}. These results could only be achieved with grid refinement. The refined grid spacing is one third of the coarse grid spacing in the radial and one fourth in the angular direction.

\section{Numerical simulation of semiconvection}

When the convective transport does not depend on the temperature gradient only but also on an additional scalar field like salt (in the ocean) or helium (in stars), then \textit{double-diffusive convection} can occur. This name implies that two diffusion coefficients ($\kappa_T$ and $\kappa_{S} / \kappa_{He}$) influence the fluids motion. In the oceans double-diffusive convection can form \textit{salt-fingers}, while the well-known Latte Macchiato layers are a result of another double-diffusive phenomenon, which in astrophysics is called \textit{semi-convection}.

In the evolution of stars semi-convection plays an important role. When cold hydrogen-rich matter is stratified above hot helium-rich matter, double-diffusive layering can occur under certain circumstances (Ledoux stable, $\nabla_{\mu} > \nabla-\nabla_{ad}$). The main question, which arises is the influence of the double-diffusive mixing processes on the evolution of high mass stars ($M>15 M_{\astrosun}$) around main sequence turnoff.

A stable layered structure in semi-convection zones (cf. Latte Macchiato), separated by diffusive boundary layers, is assumed  (\cite{hup76}, \cite{spruit92}). In particular, the mass and heat transport through these diffusive boundaries are of major interest for explaining mixing time scales and the life span of the entire semi-convective region. There are several unknowns when describing single layers. Especially, the thickness of a single layer as function of thermal and saline Rayleigh number ($Ra_T$, $Ra_S$) and the superadiabatic gradient is not well understood. A comprehensive hydrodynamical model, based on \cite{spruit92} and \cite{muth99} (see also  \cite{muth95}) was derived for the double diffusive convection case to tackle these uncertainties.

In \cite{zauss10a} numerical simulations in 2D of semi-convection, based on compressible and incompressible formulations, have been performed for a wide initial parameter range of the Prandtl number $\sigma=\nu/\kappa_T$, the Lewis number $\tau=\kappa_S/\kappa_T$, the modified Rayleigh number $Ra_*=Ra_T \, \sigma$ and the stability parameter $R_{\rho}=Ra_S/Ra_T$. It is proven there that simulations done in the Boussinesq approximation are directly comparable with fully compressible fluids as long as the layer heights are small enough ($H_P>>H$). The aim of this study was the verification and extent of validity of existing power laws in the form $Nu_T = \alpha Ra_T^{\beta}$, where $Nu_T$ is the thermal Nusselt number. An extrapolation to the stellar parameter space was done subsequently.

In \cite{zauss10} the parameter dependence of the fluxes is analysed. Single  and double layer simulations were at the focus of that study. While the diffusion associated dimensionless numbers ($\sigma$, $\tau$) can be compared directly between compressible and incompressible fluids, other physical quantities like the fluxes ($F_T$, $F_{S,He}$) or the Rayleigh numbers ($Ra_T$, $Ra_S$) need a special treatment. The thermal Nusselt number, which compares the diffusive heat flux to the total heat flux, has to be corrected by the adiabatic stratification flux $F_{ad}$, which is by definition not present in incompressible fluids modelled by the Boussinesq approximation. Neglecting this difference would lead to wrong comparisons in simulations with flat temperature gradients.

\begin{figure}[h!]
\begin{center}
\includegraphics[width=8cm]{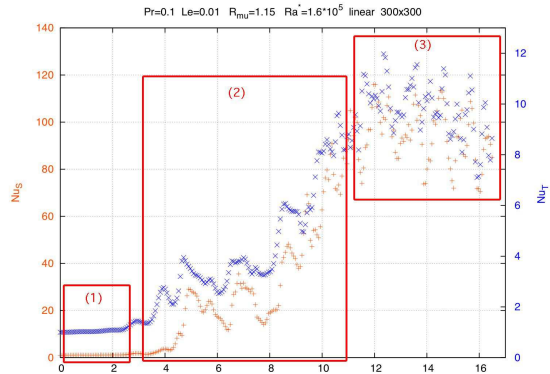}
	\caption{Nusselt numbers as function of time. The formation of a semi-convective zone (3) is the consequence of a diffusive phase (1) and
a oscillatory phase (2). For the specified parameter space the convective rolls are
established within 12 turnover cycles.} 
\label{fig:fz1}
\end{center}
\end{figure}

Linear stability analysis shows that semi-convection in an oscillatory instability. This could be observed in all our simulations. Figure \ref{fig:fz1} illustrates the development of a semiconvection cell in terms of the 
Nusselt number when starting from a linear stratification.  Nusselt numbers in the range $1<Nu_{T}<5$ indicate very diffusive regimes, where the total heat flux is dominated by conduction. High values correspond to more convective regimes. A diffusive phase (1) is followed by an oscillatory  phase (2). The fully evolved semi-convective role (3) leads to the expected (\cite{spruit92}) step like structure in helium and temperature at the upper and 
lower boundaries. With a suitable choice of parameters, a preassigned step in the middle of the 
domain is found to be stable for many eddy turnover times. 

While adoption of linear stratification as initial condition leads to very long simulation times, a `direct initial step stratification' is considered. It has been  shown that both approaches lead to the same results. Strong damping mechanisms in the parameter space $R_{\rho}>2$ and $Le>10^{-2}$ lead to very long simulation times. By starting from a step like 
stratification these could be reduced significantly and a formerly inaccessible parameter range came into reach.

\begin{figure}[h!]
\begin{center}
\includegraphics[angle=-90,width=8cm]{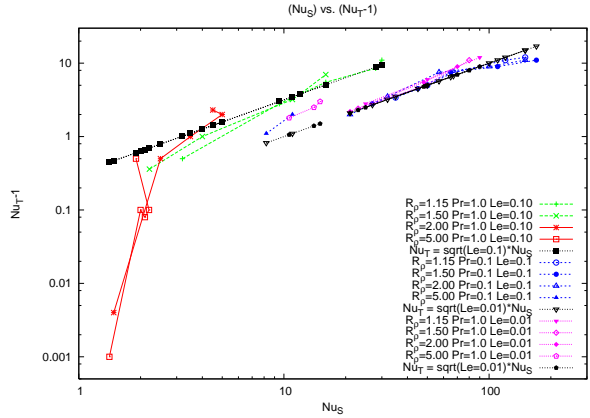}
\caption{$Nu_S$ versus $NuT-1$ plot. Black dots represent the theoretical solutions. Each line stands for a different $Ra_*$. Along each line the stability parameter decreases. The red lines denote convective sub-critial simulations, where $Ra_T<Ra_{crit}$.}
\label{fig:fz2}
\end{center}
\end{figure}

The verification of theoretical and experimental power laws under consideration for the stellar parameter space was the main focus of this study. It turned out that several existing `fitting formulas' (\cite{castaing89}, \cite{spruit92},  \cite{niemela00}) describe an upper limit. Even the flux relation,
\begin{equation}
Nu_S=\tau^{1/2}Nu_T
\label{eq:efz1}
\end{equation} 
could be verified for semi-convective stable layers (see figure \ref{fig:fz2}). Equipped with these results an extension to the stellar parameter regime was considered. The main problem of existing 1D stellar evolution codes is the lack of information about to the semi-convective mixing efficiency. The superadiabatic gradient $\nabla-\nabla_{ad}$ is not known in advance and has to be estimated. The semi-convective zone in a stellar model is limited in extent and consequently the evolution of the star on small time-scales does not depend much on the way it is calculated. But the additional diffusion in helium could be important  for later evolutionary stages. This leads to the assumption that the heat flux could be considered as known rather than the real temperature gradient $\nabla$. Based on \cite{spruit92} an implicit function was derived, relating the temperature gradient, the Rayleigh number, the fluxes and the the vertical extent of the semi-convective zone to each other.

\textbf{Acknowledgements} This research has been supported by grants P18224, P20762, P20973 and P21742 of the Austrian 
Science Foundation and by grant KU 1954/3 within SPP 1276 of the Deutsche Forschungsgemeinschaft. Calculations were performed within DEISA (project SOLEX), at the Leibniz Computer Center, Munich (project SOLARSURF), at computers of the Max Planck Society and at the VSC cluster, Vienna. We are grateful to G. Houdek, Vienna, for 
providing us with starting models for cepheids.

\begin{discussion}

\discuss{Trampedach} {Your simulation of semi-convection looked 2-D. Is that correct? Will it be extended to 3-D?}
 
\discuss{Muthsam}{Yes, the present models are 2D. We will turn to 3D in the forseeable future. }

\discuss{Gough}
{The stated purpose of the final solution you presented was to determine whether the interface between turbulent convection layer is purely diffusive, but you did not tell us whether that is the case. Please would you tell us whether the transport  of heat and helium across the interface which you showed us, which maintains its integrity despite being modulatory is purely diffusive or whether there is significant (small scale) convective transport?
}

\discuss{Muthsam}
{You obviously refer to the movie with the pre-assigned step in the middle of the 
domain. The interface itself looks quite stable. Only on isolated spots (upstreams 
or downstreams) these streams can penetrate it somewhat.}

\discuss{Pouquet}{Do you see any advantage to use a hybrid (MPI/OpenMP) parallelization method?}

\discuss{Muthsam}
{For really large runs you have to use MPI anyway for lack of shared memory. 
If you have, on one node, multiple cores you save messages when using them 
in the OpenMP rather than in the MPI mode.}

\end{discussion}

\end{document}